\documentclass[lettersize,journal]{IEEEtran}
\usepackage{amsmath,amsfonts}
\usepackage{algorithmic}
\usepackage{array}
\usepackage[caption=false,font=normalsize,labelfont=sf,textfont=sf]{subfig}
\usepackage{textcomp}
\usepackage{stfloats}
\usepackage{url}
\usepackage{amsmath}
\usepackage{makecell}
\usepackage{balance}
\usepackage{bbm}
\usepackage[linesnumbered, ruled,vlined]{algorithm2e}
\usepackage{amsmath}
\usepackage[hidelinks]{hyperref}
\usepackage{verbatim}
\usepackage{graphicx}
\usepackage{cite}
\usepackage[dvipsnames,svgnames]{xcolor}
\usepackage[most]{tcolorbox}
\usepackage{enumitem}
\usepackage{setspace}
\usepackage{float}
\usepackage{orcidlink}
\usepackage{tabu} 
\usepackage{booktabs}
\usepackage{multirow}
\usepackage[normalem]{ulem}
\useunder{\uline}{\ul}{}
\usepackage{amssymb}
\setlist{noitemsep,parsep=0pt,partopsep=0pt} 
\newcommand{\eg}{e.g.}
\newcommand{\ie}{i.e.}
\newcommand{\etal}{et al.}
\newcommand{\systemname}{\textit{\textit{CellScout}}\xspace}

\definecolor{review}{HTML}{f72585}
\usepackage{marginnote}
\usepackage{adjustbox}
\global\marginparsep=9pt
\newcommand{\sidecomment}[1]{%
  \ifdefined\revise
  \marginnote{%
    \textcolor{review}{%
      \adjustbox{minipage=1.2\marginparwidth,fbox}{%
          \scriptsize#1%
      }
    }
  }
  \fi
}

\usepackage{mathptmx}                  
\usepackage{CJK}

\definecolor{tvcghighlight}{RGB}{18, 107, 174}

\newcommand{\rui}[1]{\textcolor{black}{#1}}

\definecolor{border}{RGB}{100, 100, 100}

\newtcbox{\mybox}[1][border]
  {on line, arc = 1pt, outer arc = 1pt, colframe = border,
    colback = #1!4!white, boxsep = 0pt, left = 1pt, right = 1pt, top = 1pt, bottom = 1pt, boxrule = 1pt}

\AtBeginDocument{

}

\begin{document}
\title{\systemname: Visual Analytics for Mining Biomarkers in Cell State Discovery} 

\author{Rui Sheng\orcidlink{0000-0001-9321-6756}, Zelin Zang$^{*}$\orcidlink{0000-0003-2831-5437}, Jiachen Wang\orcidlink{0000-0001-9630-9958
}, Yan Luo\orcidlink{0009-0005-9799-3540}, Zixin Chen\orcidlink{0000-0001-6036-2351}, Yan Zhou\orcidlink{0000-0003-3581-4701}, Shaolun Ruan\orcidlink{0000-0002-6163-9786}, Huamin Qu\orcidlink{0000-0002-3344-9694}
\thanks{Rui Sheng, Yan Luo, Zixin Chen, Huamin Qu are with the Hong Kong University of Science and Technology; Zelin Zang and Yan Zhou is with Westlake University; 
Zelin Zang is also with CAIR, Hong Kong Institute of Science \& Innovation, Chinese Academy of Sciences;
Jiachen Wang is with Zhejiang University; Shaolun Ruan is with Singapore Management University;
}
\thanks{$^{*}$Corresponding author}
\thanks{The manuscript has been accepted by IEEE TVCG, and the final version is available at 10.1109/TVCG.2025.3636102.}
\thanks{© 20XX IEEE.  Personal use of this material is permitted.  Permission from IEEE must be obtained for all other uses, in any current or future media, including reprinting/republishing this material for advertising or promotional purposes, creating new collective works, for resale or redistribution to servers or lists, or reuse of any copyrighted component of this work in other works.}
}

\markboth{Journal of \LaTeX\ Class Files,~Vol.~14, No.~8, August~2026}%
{Shell \MakeLowercase{\textit{et al.}}: A Sample Article Using IEEEtran.cls for IEEE Journals}



\maketitle

\begin{abstract}
Cell state discovery is crucial for understanding biological systems and enhancing medical outcomes. A key aspect of this process is identifying distinct biomarkers that define specific cell states. However, difficulties arise from the co-discovery process of cell states and biomarkers: 
\rui{biologists often use dimensionality reduction to visualize cells in a two-dimensional space. Then they usually interpret visually clustered cells as distinct states, from which they seek to identify unique biomarkers.} 
\sidecomment{R1C2}
However, this assumption is often invalid due to internal inconsistencies in a cluster, making the process trial-and-error and highly uncertain. Therefore, biologists urgently need effective tools to help uncover the hidden association relationships between different cell populations and their potential biomarkers. To address this problem, we first designed a machine-learning algorithm based on the Mixture-of-Experts (MoE) technique to identify meaningful associations between cell populations and biomarkers. 
\sidecomment{R2C4}
\rui{We further developed a visual analytics system--\systemname--in collaboration with biologists}, to help them explore and refine these association relationships to advance cell state discovery. 
We validated our system through expert interviews, from which we further selected a representative case to demonstrate its effectiveness in discovering new cell states.
\end{abstract}

\begin{IEEEkeywords}
Biomedical, Cell State Discovery, Gene Expression Data, Machine Learning, Human-AI Collaboration
\end{IEEEkeywords}

\section{Introduction}
\sidecomment{R1C1}
\rui{Understanding how cells transition between different functional states is fundamental to explaining biological processes and disease mechanisms. However, our current knowledge of cell states remains incomplete and continuously evolving, as new states continue to emerge in response to developmental, environmental, and pathological contexts.}
Cell state discovery thus plays a critical role in advancing our understanding of biological systems and improving medical outcomes~\cite{jindal2018discovery, ong2015classification, Jonas2015Automatic, Carlo2012Mechanical}. 
By characterizing distinct cell states, biologists can uncover the mechanisms underlying various diseases such as cancer and autoimmune conditions~\cite{craig2024mmil, Aguirre-Gamboa2013SurvExpress}, which further serves personalized medicine.
One of the significant aspects of discovering cell states is to identify distinct biomarkers for each state from large-scale gene expression data. 
Specifically, these biomarkers are characterized by the unique expression of particular genes within a cell group.
For instance, in dendritic cells, the high expression levels of CD80 and CD86 genes serve as an important biomarker in distinguishing the activation state of dendritic cells.

Identifying biomarkers for various cell states presents a significant challenge for biologists. Most existing automated algorithms are designed to identify biomarkers for known cell states~\cite{Nicolas2015Cancer, Rajpal2023XAI, Zhou2024BAMBI, Reitsam2024Artificial}. However, when a cell state is undefined, the identification of biomarkers and the determination of cell states are often regarded as a co-discovery process. In this context, biologists typically select specific cell populations based on visually clustered cells in a two-dimensional space, assuming this clustered population represents a distinct cell state. Then, they compare them with other cells to identify distinct biomarkers. Unfortunately, this assumption is often invalid since the selected population may not accurately reflect a unique cell state, as evidenced by the inability to identify meaningful biomarkers to characterize those selected cells. Therefore, there is a critical need for effective tools to enhance biologists' understanding of the inherent association relationships between cell populations and potential biomarkers.

Addressing this problem is challenging. Designing an automatic algorithm to extract meaningful associations between biomarkers and cell populations is difficult, as biologists can have various exploration objectives—some explore cell states from a disease perspective, while others might focus on discovering new states based on cell development. This leads to varying definitions of meaningful associations.
Additionally, their objectives are sometimes heuristic and emerge during the exploration process, making it difficult to ascertain their priors. Consequently, how to effectively mine meaningful association relationships from a vast array of possibilities remains unknown.
Second, even after mining some potential association relationships, biologists still face challenges in interpreting, constructing, and refining cell populations and biomarkers based on the mined results.
Although various visualization tools exist, few are specifically tailored for cell state discovery. The intricate association relationships between biomarkers and cell populations require effective visualization strategies.

\sidecomment{R2C4}
\rui{To address these challenges, we developed a visual analytics system--\systemname--in collaboration with domain experts, which can help them discover new cell states and mine their biomarkers.}
To tackle the first challenge, we designed a machine-learning algorithm based on the Mixture-of-Experts (MoE) technique, 
\sidecomment{R1C1}
\rui{which refers to training multiple specialized models designed to capture distinct aspects of the input data.}
Our algorithm can mine multiple association relationships that maximize the information available to experts, which ensures that it provides valuable assistance in their exploration even when experts' priors are unclear.
To tackle the second challenge, we created an interface to help biologists understand and refine the results produced by our model.
Specifically, we feature a novel design to help experts better associate biomarkers with cell populations.
Our contribution can be summarized as follows:
\begin{itemize}
    \item We formulate the design requirements of the visual analytics system by interviewing four experts to effectively support cell state discovery and biomarker mining.
    \item We include a novel MoE-based algorithm to mine meaningful associations between cell populations and potential biomarkers in the absence of clear exploration objectives.
    \item We create a visual analytics interface featuring a novel design to help biologists understand and refine the results produced by our model.
    \item We validate our system's usability and effectiveness through expert interviews and demonstrate its utility in discovering new cell states with a real-world dataset through a case.
    
\end{itemize}
\section{Related Work}
\subsection{Cell State Discovery}
Cell states refer to the specific functional and phenotypic status of a cell at a given time.
Identifying cell states can provide insights into disease mechanisms and potential drug targets~\cite{ craig2024mmil, Aguirre-Gamboa2013SurvExpress}, which is significantly important for the medical domain.
The most common method for identifying a unique cell state is to find specific biomarkers through single-cell gene expression datasets~\cite{LINDEBOOM2021625, Dann2022Precise}.
Specifically, unsupervised clustering approaches have been widely used in cell state discovery~\cite{trapnell2015defining, Trygve2017Cell, Aevermann2018Cell, Aevermann2021machine}.
This method first leverages clustering algorithms to group cells based on their representation. Then each group is assumed to represent a distinct cell state, and statistical or machine-learning methods are employed to identify biomarkers.
However, these methods have significant shortcomings. They fail to ensure that the resulting clusters accurately reflect biologically distinct groups associated with specific cell states~\cite{miao2020putative}.
Therefore, a considerable amount of manual labor is often required for result refinement and adjustments.
Miao \etal~\cite{miao2020putative} proposed a new method that iteratively applies a machine-learning approach to merge clusters that may represent the same cell state, significantly reducing the need for manual intervention.
However, this method still falls short of capturing the internal complexities of each cluster, as there may be notable differences within a single cluster.
Therefore, there is a need to design a method that can reveal more fine-grained association relationships between biomarkers and different cell populations, which can facilitate the cell state discovery process.

\subsection{Visual Analytics for Gene-related Data Analysis}
Visual analytics has been widely used for analyzing gene-related data in the biomedical area~\cite{schroeder2013visualizing, cerami2012cbio, Krueger2020Facetto, Somarakis2021ImaCytE, hong2022lineaged, Hong2024CellLineage, Sehi2023Multi-View, Sehi2022Cosling, Smits2024Explaining, Halladjian2022Multiscale, Brandt2025Understanding, Nusrat2019Task}.
Some authors~\cite{van2021genova, pandey2022genorec, Sehi2023Multi-View} developed visual analytics tools to help biologists visualize and examine high-throughput genomic data.
For example, GenoREC~\cite{pandey2022genorec} is a visualization recommendation system designed specifically for genomics data that can help analysts choose effective visualization techniques based on their data characteristics and analysis tasks. 
L'Yi \etal~\cite{Sehi2023Multi-View} explored various genomics visualization tools available on the web and introduced a flexible framework that can be applied to create interactive, multi-dimensional views for genomics data analysis.
\sidecomment{R1C3, R3C1, R3C2}
\rui{
Several other studies focus on helping biologists complete specific tasks, such as alternative splicing~\cite{Fernandes2024Geneapp, Strobelt2016Vials}, cell lineage construction~\cite{hong2022lineaged, Hong2024CellLineage}, and differential gene expression analysis~\cite{Basu2025CytosploreEvoViewer, Soumyadeep2023CSV}. 
For example, Cytosplore EvoViewer~\cite{Basu2025CytosploreEvoViewer} enables researchers to visually explore and relate transcriptomic data from different species.
However, they are not designed for discovering previously unknown cell states.
The most similar works use visualization to identify cell types or phenotypes~\cite{Krueger2020Facetto, Dinkla2017Screenit, Schapiro2017histoCAT, hollt2016cytosplore}. Though they focus on detecting broad cell categories rather than capturing finer-grained processes such as functional cell states, they follow the same underlying discovery process as cell state discovery.
However, these approaches still depend on experts to infer cell types or phenotypes from visually identified clusters in two-dimensional projections. As a result, they struggle to help experts recognize cases where cells within the same cluster may actually belong to different states, or where cells in different clusters may share the same state.}

\subsection{Visual Analytics for Association Rule Mining}
Association rule mining is a crucial data mining technique aimed at uncovering intriguing relationships and patterns within large datasets~\cite{kaushik2021systematic, Meera2021Discovering, Bertl2023Finding}.
Visualizing the discovered association rules is a crucial component of the association rule mining process. It helps users better comprehend the outcomes of the rule mining, thereby enhancing their understanding of the results~\cite{Iztok2023Survey}.
For example, multiple works have designed various visualization methods to present large numbers of association rules, such as Matrix~\cite{hahsler2017visualizing, ong2002crystalclear}.
Additionally, association rules have also been used in multiple scenarios.
For example, RuleMatrix~\cite{Ming2019RuleMatrix} leveraged mined association rules as an interpretability mechanism in machine-learning models.
However, unlike traditional association rule mining, where all attributes in the dataset have defined labels, the labels of cell states in our association relationships are unknown. Therefore, we must consider what types of associations to mine that can assist experts in better understanding the inherent relationships within the dataset, ultimately aiding in the cell state discovery process. 
This distinction necessitates careful algorithm design and the support of new visualization methods.

\section{Methodology}
We collaborated closely with four biologists (E1-E4) who specialize in cell state discovery over a six-month period for our project. 
\sidecomment{R2C3, R2C4}
\rui{Through interviews, we defined the problem and identified the design requirements. Our collaboration also involved regular meetings and iterative feedback loops to refine the system design.} E1 and E2, both co-authors of our paper, have 5 and 4 years of experience, respectively. E3 and E4 are external experts with 5 and 8 years of experience.
We held weekly meetings with E1 and E2 to ensure that we were making consistent progress and incorporating their feedback in real time, allowing for fast iterations and adjustments to our approach.
For E3 and E4, our meetings occurred every two weeks, to supplement the requirements provided by the other two domain experts.
Additionally, we obtained IRB approval.

\subsection{Problem Formulation}
\begin{figure}[ht]
\centering
\vspace{-1em}
\includegraphics[width=\linewidth]{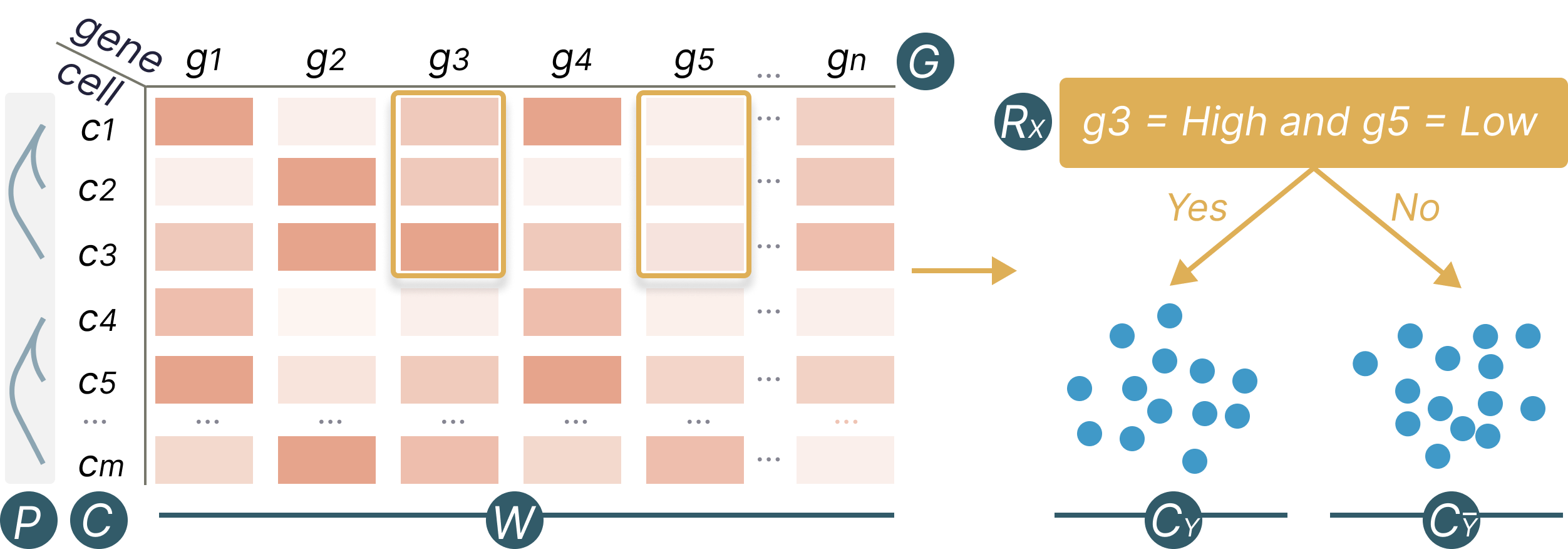} 
\caption{This figure illustrates the overview of the problem formulation. Here, \( P \) represents cell representation and can indicate the similarity among different cells based on their gene expression levels. Then, \( C \) denotes individual cells, while \( G \) refers to different gene types. \( W \) is a matrix composed of \( C \) and \( G \), representing the expression levels of various genes in different cells. Based on this data and the user's domain knowledge \( D \), experts aim to extract different association relationships \( R \), each of which can be used to distinguish a group of cells \( C_Y \) based on the uniqueness of their gene expression.}
\label{fig: problem}
\end{figure}

Our single-cell gene expression dataset can be represented as a matrix \( W \in \mathbb{R}^{m \times n} \), where \( m \) denotes the number of individual cells and \( n \) represents the number of genes measured in each cell. 
Therefore, \( C = \{c_{1}, c_{2}, \ldots, c_{m}\} \) denotes the individual cells, while \( G = \{g_{1}, g_{2}, \ldots, g_{n}\} \) represents the types of genes.
\sidecomment{R3C3}
\rui{Usually, the number of cells can reach into the thousands, with the involved genes numbering in the hundreds.}
The entry \( W_{ij} \) indicates the expression level of a gene \( g_j \) in cell \( c_i \).
This expression level reflects the amount of mRNA produced from the gene in that specific cell, which is typically quantified as a numerical value.
A higher value indicates a higher expression level of the gene. In general, biologists are more concerned with the relative expression levels of a gene across different cells, rather than its absolute value, as the comparison between cells is more meaningful.
To better assist in understanding our problem definition, we provide an illustration below (\autoref{fig: problem}).

The goal of domain experts is to identify association relationships \( R(X) \Rightarrow Y \) while considering their own domain knowledge \( D \) and cell representation \( P \). 
The concrete explanation of each parameter in \( R(X) \Rightarrow Y \) is as follows:
\begin{itemize}
    \item \( X \) is a subset composed of attributes from \( G \), which can be represented as \(X = \{(g_i) \mid g_i \in G \}\).
    \item \( Y \) is the state label of a cell subset \( C_Y = \{ (c_i) \mid c_i \in C \} \). \( Y \) is uncertain and can be determined based on expert knowledge \( D \) during the exploration process. Furthermore, the cell representation \( P \) imposes certain constraints on the composition of \( C_Y \), indicating that those represented closely are more likely to share the same state.
    \item \( R \) signifies that the presence of \( X \) indicates that the cell population belongs to state \( Y \). This also implies that \( X \) does not infer the presence of any alternative states \( Y' \). Specifically, \( R \) can be represented as constraints on \( X \): \(X \rightarrow \{(g, r) \mid g \in G_X \}\), where \( g \) represents a gene and \( r \) denotes its range of expression levels.
\end{itemize}
Our mining method aims to identify multiple meaningful association relationships \( R \) under the constraint of \( P \) to assist experts in discovering cell states, even when their prior knowledge \( D \) is uncertain.

\subsection{Design Requirements}
\sidecomment{R2C6, R2C8}
\rui{Even with our MoE-based mining algorithm, the final identification of cell states and biomarkers cannot be fully automated, requiring expert exploration. In this process, experts face two main challenges: first, interpreting association relationships between genes and cell populations uncovered by the MoE model; and second, refining and modifying these results based on their domain knowledge, which may include adjusting, combining, or reconstructing associations to obtain biologically meaningful insights.
Therefore, we conducted a thematic analysis of the interview transcripts collected from regular meetings with our experts at the early stage of this project to identify common design requirements. Three authors independently coded the transcripts, then discussed their findings to reach a consensus. This process resulted in five design requirements that describe how biologists aim to utilize the mined association relationships.}

\begin{itemize}[leftmargin=1.8em]
\item[\textbf{T1}] \textbf{Examine mined association relationships} between cell populations and potential biomarkers.
Experts need to review the mined association relationships and identify the ones they are interested in.
This involves two main points: first, providing experts with an overview of how different association relationships are linked to specific regions of cell populations. Second, it should highlight the genes that might constitute the corresponding biomarkers. 

\item[\textbf{T2}] \textbf{Explore cell representations} with the assistance of mined associations relationships.
Experts often examine the similarity in gene expression between cells: the more similar their expressions, the more likely they belong to the same cell state. 
Therefore, we should provide a clear visualization of cell representations. Given that each cell has hundreds or thousands of dimensions of gene types, domain experts usually rely on dimensionality reduction methods (\eg, PCA) to illustrate the similarity between cells, ultimately leading to the generation of two-dimensional cell representations.
Additionally, we should consider how to integrate our mined results into the cell representation exploration process. For instance, E2 mentioned the desire for \textit{``insights from the mined results''}. E3 also noted, \textit{``The extracted association relationships are intermediate results that might be helpful and  that I hope to use for analyzing and exploring gene-level data.''}

\item[\textbf{T3}] \textbf{Compare differences} among different cell populations.
Experts may be interested in several cell populations while exploring cell representations. Therefore, they need to compare gene types and identify differences in their expression levels across these areas of interest.
For example, E5 mentioned, \textit{``If I find that two clusters of cell populations are very close to each other, I might be uncertain whether they actually belong to the same cell state. At that point, I would like to quickly compare them.''}
In addition, understanding which gene types show differential expression levels between the two regions can help domain experts better identify the composition of potential biomarkers.

\item[\textbf{T4}] \textbf{Modify and refine} cell populations and the associated biomarkers. 
Our automatically mined results may not fully meet the experts' needs. In such cases, experts may need to use the mined association relationships to reconstruct the potential biomarkers for their cell populations of interest. Additionally, E2 mentioned, \textit{``We will place great importance on whether the genes that constitute a biomarker are biologically meaningful together; this may require judgment and modification from the ourselves.''} 
Therefore, we should provide a flexible way to help refine the mined results based on 
\sidecomment{R1C4\_2}
experts' domain knowledge, 
\rui{ensuring that the selected genes are biologically meaningful (i.e., that these genes collectively can be reasonably interpreted by the experts from a biological perspective, such as disease).}

\item[\textbf{T5}] \textbf{Support statistical tests} for the identified biomarkers.
During the refinement of biomarkers, experts need to know how accurately the identified biomarkers can distinguish the corresponding cell populations from others. To achieve this, our system should provide metrics for performance measurement. Specifically, E4 pointed out that a commonly used metric is the F1-score; in addition, E2 and E5 also suggested including accuracy as a metric.
\end{itemize}
\section{System Design}
\begin{figure}[ht]
\centering
\includegraphics[width=\linewidth]{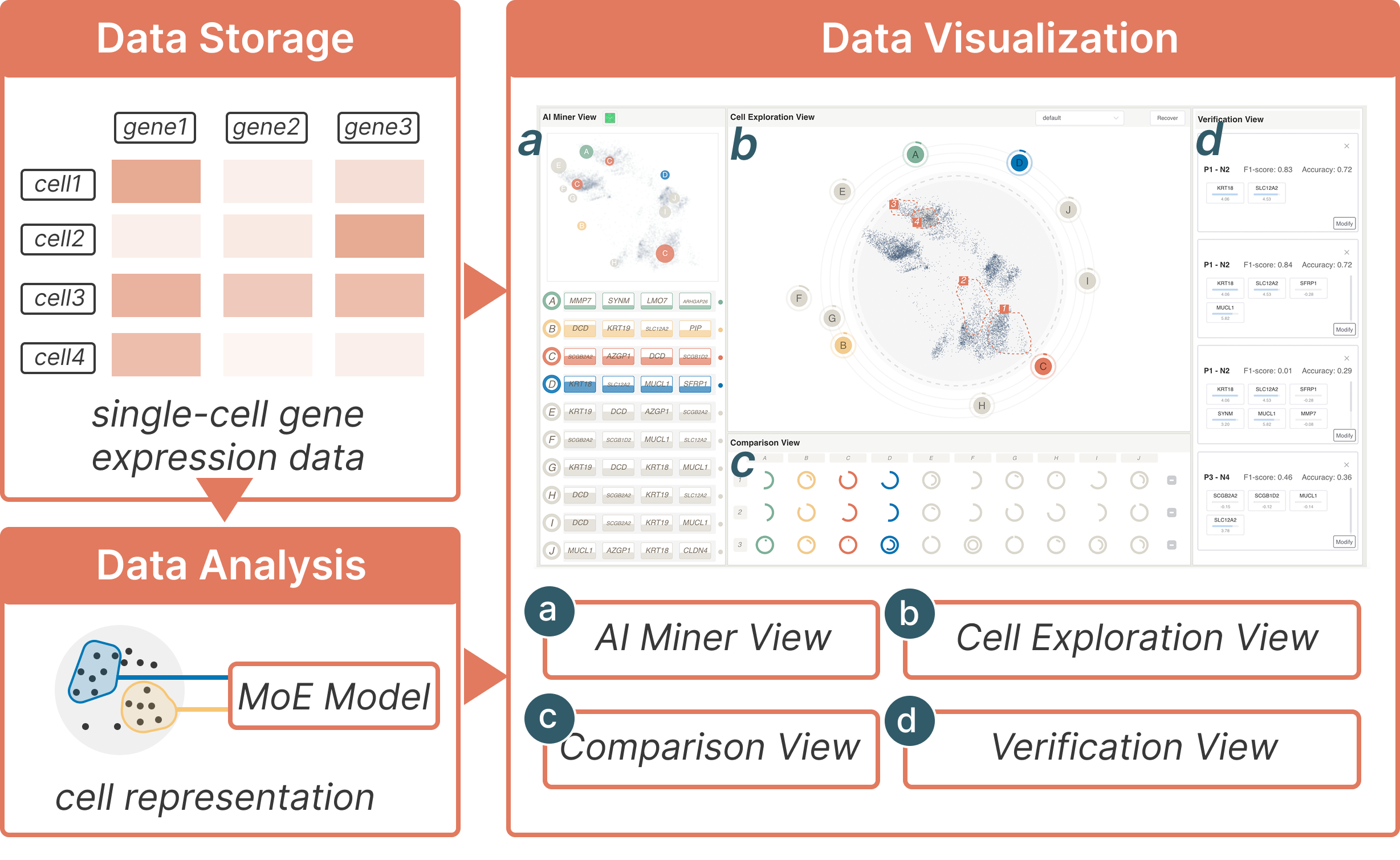}  
\caption{The system overview of \systemname. There are three components: the Data Storage Component, the Data Analysis Component, and the Data Visualization Component.}
\label{fig: framework}
\end{figure}

Our system has three main components (\autoref{fig: framework}): the Data Storage Component for storing single-cell gene expression data, the Data Analysis Component for mining associations between cell populations and potential biomarkers, and the Data Visualization Component for presenting gene expression data and mined relationships to help cell state discovery. 
We will first introduce our model section, and then we will explain how we complete the visualization design to enable experts to effectively use our model.

\subsection{Mining Model}\label{model}
To extract meaningful associations between potential biomarkers \( \mathcal{B} \) and uncertain cell states \( Y \), we introduce an optimization framework that accounts for the uncertainty inherent in expert domain knowledge \( D \) and the constraints of cell representation \( P \).
Specifically, our goal is to ensure that the association relationships we uncover encompass the maximum amount of information about the inherent associations in the gene expression dataset while being sufficiently distinct from one another. 
Through this approach, our extracted association relationships can inform experts, even in the context of ambiguous exploration objectives \( D \).
To achieve this, we have designed our mining method based on the Mixture-of-Experts (MoE) framework.

\subsubsection{Mixture-of-Experts Model}
The MoE model is a machine learning strategy that primarily involves partitioning the input data into distinct clusters or subsets, 
\sidecomment{R1C4\_4}
\rui{with each AI expert—an independently trained neural network—specializing in a specific portion of the data~\cite{masoudnia2014mixture}}. Multiple works~\cite{zang2024dmthi, Yuksel2010Variational, Yan2024Efficient, CHEN2024112056} have leveraged the MoE approach to capture diverse patterns within the data.
Specifically, the MoE model first consists of \( k \) AI experts (\ie, $f_1, f_2, \ldots, f_k$), each with its unique parameters that need to be trained.
Typically, the value of $k$ can be determined through testing, as increasing $k$ may lead to stable performance and, in some cases, even a decline in performance.
Each AI expert will take the same input \( x \) and generate outputs $f_1(x), f_2(x), \ldots, f_k(x)$.
Additionally, the MoE framework includes a gating mechanism to select the appropriate subset for each AI expert, where the function \( w \) takes the input \( x \) and generates a vector of outputs:
\[
w_1(x), w_2(x), \ldots, w_k(x).
\]
Therefore, given an input \( x \), the MoE model generates an output by combining the results from the AI experts \( f_1(x), \ldots, f_k(x) \) with corresponding weights \( w_1(x), \ldots, w_k(x) \). The output is computed as:
\[
f(x) = \sum_{u} w_{u}(x) f_{u}(x).
\]
We have the flexibility to define different loss functions tailored to specific tasks, such as cross-entropy for classification problems or mean squared error for regression tasks.

\subsubsection{Designed Method}
To address our specific domain problem, we designed a novel MoE model that can be used to mine association relationships between cell populations and potential biomarkers.
Building upon DMT-HI~\cite{zang2024dmthi}, a general-purpose dimensionality reduction method, our model is further tailored to the cell state discovery scenario by incorporating domain-specific constraints and optimization objectives.
Specifically, in our MoE mining model, different AI experts are designed to capture distinct \textit{``cell population–gene subset''} associations.
Their extracted results can reveal the relevance between each cell and a given association, as well as the importance of each gene within that association.
\sidecomment{R2C2}
\rui{For each cell, the model computes a relevance score indicating how well the cell matches a given association relationship.
In addition, the model evaluates the discriminative power of each gene in distinguishing a given association relationship from others. Genes with higher scores contribute more to distinguishing association relationships and are regarded as more important biomarker candidates.}
We have provided the pseudocode (Algorithm 1) to enhance clarity. We then present a detailed explanation of our approach.

Formally, the input to our model is the gene expression matrix $W$.
The sets of cells and genes are $C$ and $G$, respectively.
Then, we define a mapping from the original high-dimensional gene expression space to a biomarker subspace as:
$
\mathcal{M}: W \rightarrow \mathcal{B},
$
where \( \mathcal{B} = \{ b \mid b \subseteq G \} \) represents the selected biomarkers. 
Our optimization objective of the MoE model should first maximize the discriminative power of selected biomarkers. Therefore, we define the loss function as follows:
\[
\max_{\mathcal{B}} \mathbb{E}_{C, Y} \left[ F(\mathcal{B}) \right], \quad \text{with} \quad F(\mathcal{B}) = \sum_{b \in \mathcal{B}} p(b \mid Y) \log\frac{1}{p(b)},
\]
where \( p(b \mid Y) \) is the conditional probability distribution of biomarker \( b \) across distinct cell populations \( C_{Y} \).
Then, our approach should aim to maximize the amount of information provided by the association relationships as much as possible. To evaluate this, we can assess how well we preserve the informational content of the original gene expression data. Specifically, we can leverage a metric known as mutual information retention (MIR), which is represented as:
\[
I(G, \mathcal{B}) = \frac{H(G) + H(\mathcal{B}) - H(G, \mathcal{B})}{H(G)},
\]
where \( H(\cdot) \) denotes the entropy. MIR quantifies how much information from the original gene expression space \( G \) is retained within the biomarker subspace \( \mathcal{B} \).
Crucially, considering the constraint of cell representation $P$, which indicates that cells with similar representations are more likely to share the same state, we introduce a cell representation constraint (CRC). CRC ensures that closely represented cells in the embedding space are more likely to share similar cell states:
\[
    \| e(c_i) - e(c_j) \|^2 \leq \delta \Rightarrow p(Y_i = Y_j) \geq \gamma,
\]
where $e(c_i)$ and $e(c_j)$ represent the embeddings of two distinct cells, and \(\delta\) and \(\gamma\) are predefined hyperparameters controlling representational similarity.
Specifically, we calculated the average distance of 5 neighborhoods of each center point as \(\delta\) and correspondingly set \(\gamma\) to 0.1.
This design choice was primarily motivated by our desire to avoid making the constraint too strict, which could lead the model to overlook other more important optimization objectives.
This constraint implies that the more similar the embeddings of $c_i$ and $c_j$, the greater the probability that they belong to the same cell state.
Thus, our overall optimization objective can be presented as:
\[
    \max_{\mathcal{B}} \mathbb{E}_{C, Y}\left[ F(\mathcal{B}) \right] + \lambda I(G, \mathcal{B}), \quad \text{s.t. CRC}.
\]
The expected value \( \mathbb{E}_{C, Y}[ F(\mathcal{B})] \) lies within the range \( [0, \log |\mathcal{B}|] \), while the mutual information \( I(G, \mathcal{B}) \) is bounded between 0 and 1. To ensure a balanced comparison between these two quantities, we define the hyperparameter \( \lambda \) as \( \log |\mathcal{B}| \) in our experiment.

\begin{algorithm}[t]
\caption{MoE-based Mining Model}
\label{alg:biomarker_mining}
\KwIn{Gene expression matrix $M$, Number of AI experts $k$, Cell representation constraint $\delta$, Similarity threshold $\gamma$, Loss weight $\lambda$, Learning rate $\alpha$, Epochs $Epo$}
\KwOut{Extracted association relationships $R$}

Initialize experts $\{f_1, f_2, \ldots, f_k\}$ with parameters $W$\;

\For{epoch = 1 to $Epo$}{    
    \For{each AI expert $f_u$ in $\{f_1, \ldots, f_k\}$}{
        Apply Gumbel-Softmax gating mechanism to select gene subsets\;
    }
    $\text{Aggregate}(\{f_1, \ldots, f_k\})$\;
    
    Calculate discriminative power loss $F(\mathcal{B})$\;
    Calculate mutual information retention $I(G, \mathcal{B})$\;
    
    Apply cell representation constraint (CRC): $\|e(c_i) - e(c_j)\|^2 \leq \delta \implies p(Y_i = Y_j) \geq \gamma$\;
    
    Optimize model parameters $\{W, \text{gating weights}\}$: \(\max_{\mathcal{R}, \mathcal{M}} \mathbb{E}_{C, \mathcal{B}}\left[ F(\mathcal{R}) \right] + \lambda I(G, \mathcal{B}), \quad \text{s.t. CRC}.\);
}

Extract association relationships $R$ for each expert:\;
$R_u \gets \{ (\text{Rel}(c_p), \text{Imp}(g_q)) \mid c_p \in C, g_q \in G \}$\;

\Return{$R$}\;
\end{algorithm}


To assist each AI expert in the MoE framework in dynamically selecting gene subsets, we employ a Gumbel-Softmax gating mechanism, which is a widely utilized method for probabilistic selection in various MoE-based approaches. 
\sidecomment{R1C4\_6}
\rui{Ultimately, through our method, we can obtain \( k \) association relationships extracted by \( k \) AI experts, where \( k \) is a hyperparameter that can be automatically determined through experimental evaluation to find the optimal value.} Each association relationship \( R_s \) (where \( s = 1, 2, \ldots, k \)) can be represented in terms of its relevance with each cell $c_p$ in the cell dataset $C$ and the importance of each gene \( g_q \) within the gene type set $G$ as follows:
\[
R_j = \{ (Rel(c_p), Imp(g_q)) \mid c_p \in C, g_q \in G \}.
\]
This result presents to experts that each association relationship emphasizes which cell populations while simultaneously identifying the important genes (\ie, potential biomarkers) associated with these cell populations.
Notably, our model provides estimates of gene importance within biomarkers rather than directly specifying the constituent genes. This approach offers greater flexibility for experts with different objectives. 
They can utilize our results to understand inherent patterns between cell populations and potential biomarkers while also leveraging their exploratory needs and domain knowledge to select suitable genes and cell populations for the cell state discovery process.

\subsection{Visual Design}
\begin{figure*}[ht]
\centering
\includegraphics[width=\linewidth]{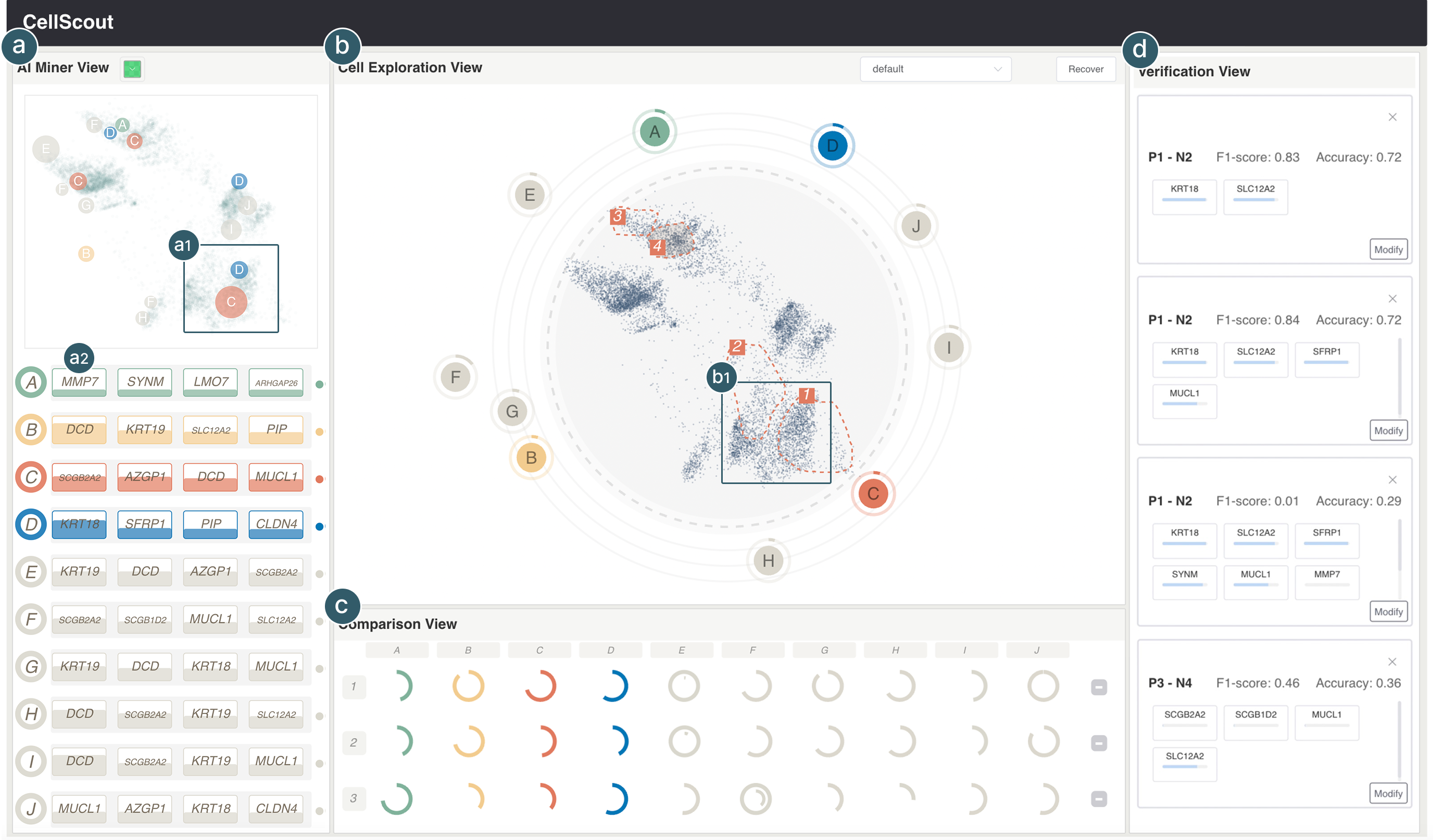}   
\caption{\textit{\systemname} integrates MoE-based techniques to identify association relationships between cell populations and their potential biomarkers, facilitating the discovery of new cell states. There are four views that, in conjunction with the mining algorithm, help biologists understand and refine mining results based on their domain knowledge. (a) In the AI Miner View, experts can access results generated by MoE-based techniques. We showcase various mined association relationships and highlight the cell populations that might be most relevant to them. Additionally, we display the importance of each gene within those relationships. (b) The Cell Exploration View helps experts effectively explore single-cell gene expression data based on their two-dimensional representations. (c) In the Comparison View, experts can compare different cell populations and assess the relevance of each association relationship. (d) In the Verification View, experts can perform statistical validation.}
\label{fig: teaser}
\end{figure*}

To help biologists better utilize our mining association relationships to identify cell states, we built a visual analytics system, \systemname. The system is composed of four views: the AI Miner View, the Cell Exploration View, the Comparison View, and the Verification View (\autoref{fig: teaser}).
Biologists can begin by examining the mining results in the AI Miner View. They can then explore cell populations according to their two-dimensional representations in the Cell Exploration View. The Comparison View supports them in comparing various cell populations based on the mined associations. Finally, the Verification View allows biologists to validate the accuracy of their discovered biomarkers and cell states through statistical verification (\eg, accuracy).

\subsubsection{AI Miner View \textbf{(T1)}}
This view (\autoref{fig: teaser}-a) visualizes the mined association relationships between potential biomarkers and cell populations.
To help experts intuitively understand the cell populations targeted by our mining model, we present a scatter plot visualization as an overview. This plot showcases the distribution of cells (represented in a two-dimensional space).
\sidecomment{R3C4}
\rui{We also highlight relatively pure regions—areas where cells share the same most relevant association relationship—by overlaying labeled dots (\autoref{fig: teaser}-a1). Specifically, to obtain these results, each cell was first labeled according to its most relevant association relationship, and DBSCAN was then applied to these labels to identify compact regions in which cells with the same label cluster together. Areas without labeled dots represent regions where cells exhibit more diverse or mixed associations.}
This design provides experts with an initial high-level overview, while more fine-grained relationship matching can be explored in the Cell Exploration View.

\begin{figure}[ht]
\centering
\includegraphics[width=\linewidth]{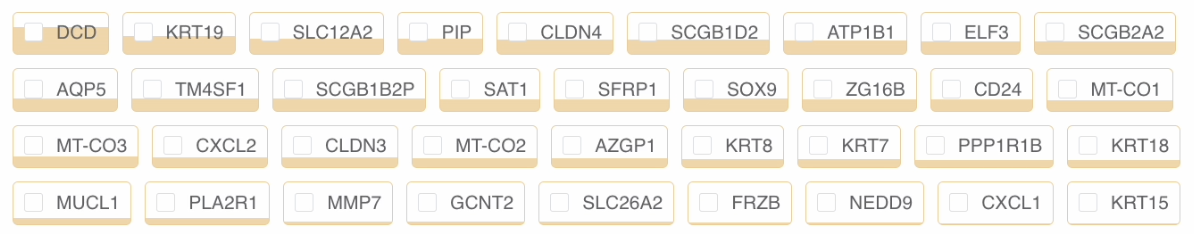}   
\caption{The gene importance scores computed by our model for each association relationship can be visualized by clicking on the corresponding association node in the AI Miner View or Cell Exploration View.}
\label{fig: importantgenes}
\end{figure}

To further help experts understand the gene composition of each association relationship, we use a bar chart to display the importance of genes within each association (\autoref{fig: teaser}-a2).
The height of the bars indicates the importance level of each gene, with taller bars suggesting a greater potential to distinguish the association relationship from others.
By default, we display the top four most relevant genes for each association relationship, as calculated by our model. However, users can click to reveal the full list of genes (~\autoref{fig: importantgenes}), ranked by their importance scores.
\sidecomment{R3C3, R3C5}
\rui{Specifically, each AI expert highlights only a small subset of key genes from its observed data, reflecting the prior biological knowledge: typically, only a few genes play crucial roles within each population. As the importance scores drop sharply, experts usually focus on a few genes, ensuring the scalability of the visualization.}

\textbf{Interaction.} Considering that our association relationships are derived through optimized hyperparameter methods, the number of such relationships may vary unpredictably. To better assist experts in distinguishing these relationships visually, we allow them to select appropriate colors for each association relationship. This approach ensures scalability in color usage, accommodating an expanding number of relationships while also providing flexibility and personalization for users.
Moreover, this approach supports dynamic exploration by allowing users to adjust color assignments in real time as their focus shifts during analysis.
\sidecomment{R3C10}
\rui{Finally, users can double-click each association to add annotations, such as their interpretation of the association or any insights they discover. The annotations are hidden by default and can be viewed via a pop-up, helping track observations, support reasoning, and allowing users to reference them later for analysis.}

\begin{figure}[ht]
\centering
\includegraphics[width=\linewidth]{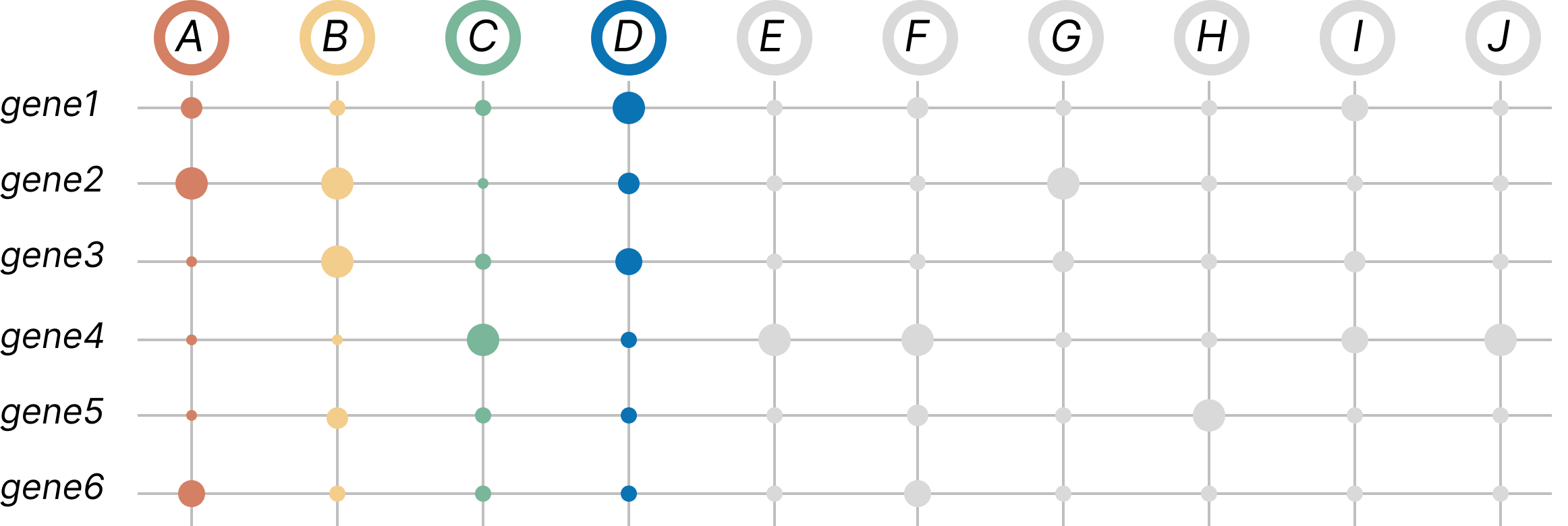}   
\caption{The alternative design represents gene importance within each mined association relationship using a matrix visualization.}
\label{fig: alternative_1}
\end{figure}

\textbf{Justification.} 
Initially, we used a matrix format (\autoref{fig: alternative_1}), which made it easy to compare the importance of a gene across different mined association relationships. However, during the iterative design process, we discovered that experts were not particularly concerned with the differences in a gene's importance across multiple mined results. For example, E1 stated, \textit{``I just need to know which genes are most important in this particular result.''} 
This feedback highlighted that experts are less focused on comparing a gene's importance across multiple relationships and more interested in identifying the key genes within a specific relationship. Consequently, we shifted to a simpler and more focused bar chart representation. Additionally, we provided an interactive feature for experts to explore the importance of other genes if needed.

\subsubsection{Cell Exploration View \textbf{(T2)}}
\sidecomment{R3C6}
\rui{This view (\autoref{fig: teaser}-b) serves as the primary component of our system, enabling experts to explore individual cells in a two-dimensional projection obtained through dimensionality reduction. 
Specifically, domain experts often rely on the spatial distances between cells in this plot to infer whether certain cells share the same state, typically assuming that cells positioned closer together are more similar. However, the two-dimensional positions of cells cannot fully guarantee the preservation of true high-dimensional relationships. To address this limitation, our mined association relationships provide additional knowledge that can enrich and guide the exploration process.} 
Therefore, this view is designed to integrate the mined results with the two-dimensional cell representations, supporting a more comprehensive analysis. 
Specifically, this view is structured to facilitate experts' exploration across three interconnected layers of data: cell representations, association relationships, and the original gene expression data.

\textbf{Cell representations.} 
For the three types of data (\ie, cell representations, association relationships, and the original gene expression data), we began by determining the visualization format for the cell representation data. 
This choice was driven by two primary considerations.
First, the cell representation data serves as the most foundational and informative component in our system, as biologists typically rely on it to gain an overview of cellular heterogeneity and to identify distinct cell states. 
Second, biologists often have well-established practices for exploring this type of data. Through our formative studies and literature survey, we found that most domain experts are accustomed to working with scatter plots based on dimensionally reduced embeddings (e.g., via t-SNE or UMAP). These scatter plots provide an intuitive and widely accepted way to examine the spatial distribution of cells and detect subpopulations. Consequently, we adopted a two-dimensional scatter plot to represent the dimensionally reduced cell representations.
\sidecomment{R3C3}
\rui{When the number of cells is large, visual clutter might obscure the detection of cell clusters. To mitigate this issue, we adjusted the transparency of each cell point in the scatter plot, enabling experts to better perceive clustering results even in cases of overlapping data.}

\begin{figure}[ht]
\centering
\includegraphics[width=\linewidth]{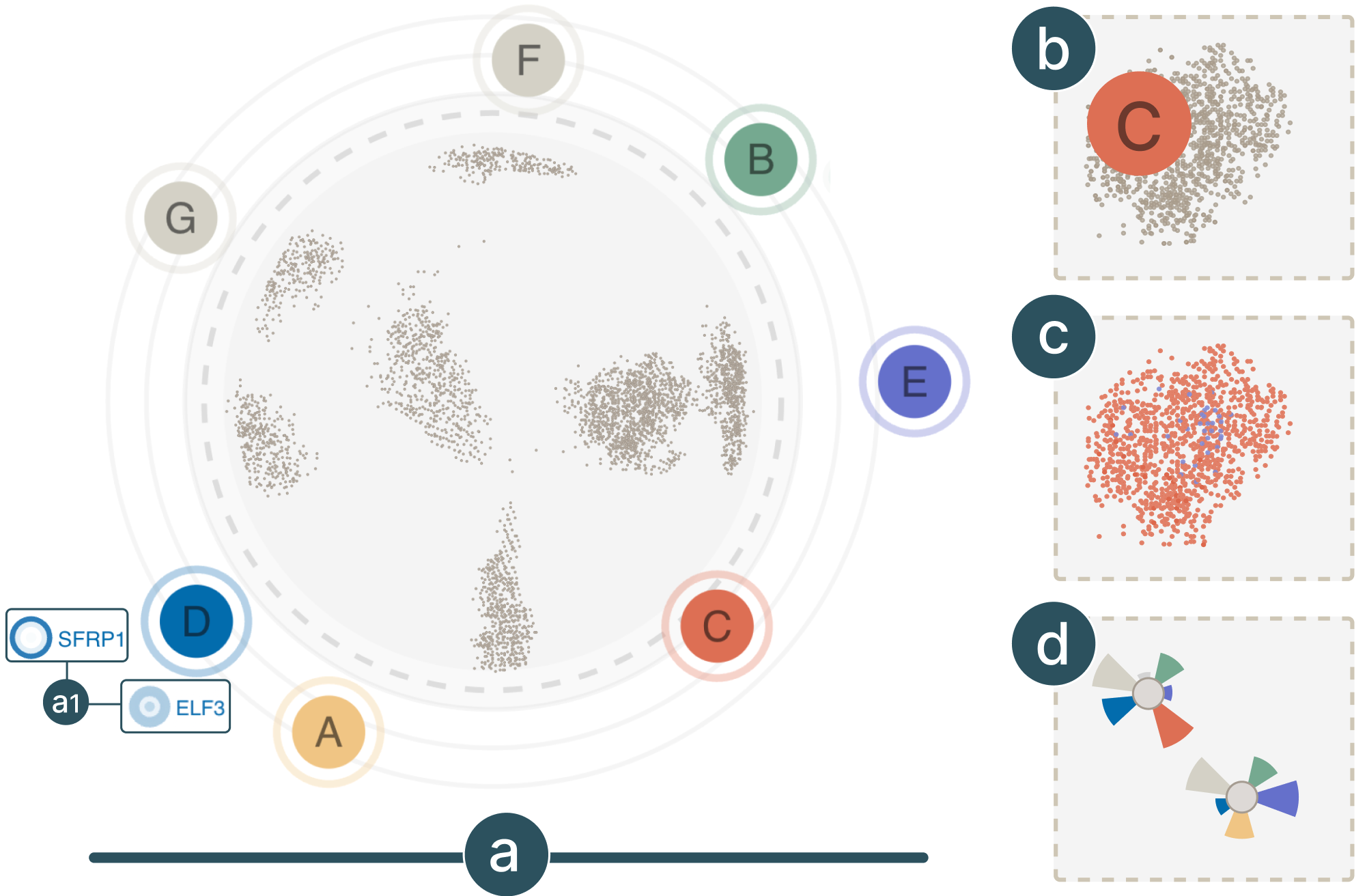}   
\caption{(a) The Cell Exploration View consists of three informational components: the cell representation data, each association relationship, and the gene expression distribution. (b-c) The alternative designs.}
\label{fig: maindesign}
\end{figure}

Our system visualizes cell representations in a polar coordinate system (\autoref{fig: maindesign}-a). This design is directly motivated by the structure of our model—a deep manifold embedding based on a MoE architecture. In this model, multiple expert subnetworks capture distinct local patterns in the cell data, and the final embedding reflects a composition of these expert perspectives.
As a result of this architecture, cells that lack strong type-specific gene expression and exhibit similarity to multiple cells tend to be positioned near the center of the embedded space. In contrast, cells with more distinctive and specialized expression profiles are pushed toward the periphery. This leads to a natural radial distribution of cells in the embedding space, where the radial distance correlates with the degree of cell-type specificity or distinctiveness. The polar coordinate system provides a direct and intuitive way to reflect this structure.
We also offer multiple dimensionality reduction methods for experts to choose from. For instance, the embeddings generated during our mining process can be used as one dimensionality reduction approach. We also include traditional methods such as t-SNE and UMAP, providing biologists with flexible options to tailor their analysis.

\textbf{Association relationships.}
To facilitate the exploration of cell representation data, we integrate mined association relationships into the Cell Exploration View. Our primary goal is to enable experts to interpret how different association relationships manifest across distinct regions of the cell embedding.
Our initial approach employed static visualization by directly modifying the original scatterplot (\ie, cell representation data), with the aim of revealing association relationships through additional encodings or glyph designs. Specifically, we first explored overlaying the corresponding association relationships directly onto the spatial regions where they were relevant (\autoref{fig: maindesign}-b). However, this led to one major issue: the overlays interfered with users’ exploration of the underlying cell representations, making it difficult to perceive the overall structure.
To address this, we encoded each cell using the color of its most relevant association relationship (\autoref{fig: maindesign}-c). While this reduced visual interference, it introduced potential bias: Cells with similar relevance to multiple associations were forced into a single color, which could oversimplify their roles and lead to misinterpretation.
Therefore, we next experimented with rose diagrams placed at each cell location (\autoref{fig: maindesign}-d), enabling a fuller representation of the association strength distribution. However, we observed that this design would quickly become difficult to interpret as the number of cells increased, suffering from severe scalability issues due to overlapping and visual congestion.

In response, we shifted toward an interaction-driven design. 
Rather than statically encoding all association relationships within the scatter plot, we enable experts to explore them on demand through interactive operations such as hovering and clicking. 
To avoid interfering with the exploration of cell representation, these association relationships are positioned outside the embedding visualization (\autoref{fig: maindesign}-a), following a solar-system-inspired layout.
Each is represented as a labeled, color-coded association node, arranged like planets in orbit.
When experts hover over an association node, the corresponding cell populations within the scatter plot are dynamically highlighted, allowing them to immediately perceive the spatial distribution of relevance.
This design ensures that we do not overly focus on only the most relevant association relationships, thereby avoiding oversimplification of the information, as seen in \autoref{fig: maindesign}-c. Moreover, it avoids the visual clutter that can arise in designs such as our rose diagrams (\autoref{fig: maindesign}-d).

To further support interpretation, we incorporate position encoding to provide a rough spatial cue about the relationship between each association and the cells it is connected to. Specifically, the angular position of an association node loosely reflects the spatial orientation of its related cell population within the scatter plot. In addition, the radial distance of an association node from the center of the scatter plot’s edge indicates how close the cells associated with that node are to the center of the embedding space—nodes linked to centrally located cells are placed nearer to the scatter plot, while those related to more peripheral cells appear farther away.
Finally, to further enhance the exploration process, each association node is surrounded by a donut chart. 
The arc length of this chart indicates the degree of relevance between the user-selected cell population and the given association relationship.

\textbf{Gene expression distribution.}
To further support the exploration, our formative study highlights the importance of providing detailed insights into gene expression distributions within selected cell populations.
In our initial design, we visualized all genes simultaneously, positioning each gene based on its computed relevance to the surrounding cell representations. While this approach offered a comprehensive view, it resulted in substantial visual clutter due to the large number of genes.
To address this issue, we analyzed expert interactions with earlier versions of the system and observed that they typically concentrated on a small set of genes. These genes were often chosen based on their importance to specific association relationships.
Guided by this insight, we redesigned the interface to enable targeted exploration of gene expression distributions. 
Experts can now explore a curated subset of genes around a selected association node, rather than viewing all genes upfront. They can either manually choose genes based on their expertise or use system-suggested genes with high importance scores (\autoref{fig: importantgenes}), accessible by clicking the association node.

Once a gene is selected, we can observe its expression distribution within the selected cell population (\autoref{fig: maindesign}-a1).
\sidecomment{R2C9}
\rui{Considering the spatial limitations and the possibility of examining multiple genes, we opted not to use the area chart, but instead developed a new design.}
Each distribution (\autoref{fig: maindesign}-a1) is presented using a circular layout, where distance from the center encodes expression levels—closer to the center indicates lower expression, and farther from the center indicates higher expression. Color intensity represents the density of cells at each expression level, with darker regions indicating a higher number of cells.
\sidecomment{R3C7}
\rui{For example, in \autoref{fig: maindesign}-a1, for the gene ELF3, there is a high density of cells with low expression, a few with medium expression, and again a high density with high expression. In contrast, for SFRP1, there are only cells expressing high levels.}
This design strikes a balance between informativeness and visual clarity.
Continuing the solar-system-inspired hierarchy, these gene visualizations are arranged like moons orbiting their association node, visually conveying their subordinate relationship.

\textbf{Interaction.}
First, users can lasso-select regions of interest within the cell population space, with each selected region automatically assigned a unique identifier and stored for future reference (\autoref{fig: teaser}-b, four regions highlighted in red).
To assess the relevance of the selected region to a specific association relationship, users can examine the surrounding visual indicators presented around the corresponding association node.
When users hover over an association node, the opacity of the cell representations dynamically updates to reflect the strength of relevance between the cells and the selected association relationship. This immediate visual feedback helps users better understand how the model-mined relationships relate to the underlying cell distributions.
Additionally, by clicking on an association node, users can view and select related genes for further exploration. The corresponding gene expression distributions are then displayed around the association node. To support personalized analysis workflows, users can manually reposition these gene expression visualizations via drag-and-drop, allowing for a more customized and organized exploration experience.

\subsubsection{Comparison View \textbf{(T3)}}
This view (\autoref{fig: teaser}-c) is primarily designed to help experts better compare differences across various cell populations by visualizing their relevance to various association relationships. 
Our model in \autoref{model} computes the relevance between each cell and every association relationship. When users select a subset of cell populations, we calculate the average relevance between this group of cells and each association. This information is then presented to facilitate comparative analysis.
To achieve this, we designed a matrix where each column represents an association relationship, and each row corresponds to a selected cell population. We utilize donut charts to depict the relevance levels, with the arc length indicating the strength of the relationship. 
\sidecomment{R1C4\_11, R3C8}
\rui{The outermost complete ring corresponds to the upper quartile of the relevance distribution across the dataset.
If the relevance exceeds the maximum value that a donut chart can represent, an additional and smaller embedded donut chart will be introduced within the original one to represent the overflow.}

\textbf{Interaction.} 
Experts can remove selected cell populations in this view if certain comparisons do not yield sufficient insights and are deemed unnecessary for further exploration. This modification will also be reflected in the Cell Exploration View, dynamically updating the visualization to ensure consistency.

\textbf{Justification.} 
Initially, we designed this view to display the distribution of all genes across different regions using area charts or heatmaps---formats that aligned with expert habits. However, during testing, we found that this approach was ineffective due to the overwhelming complexity of a large number of genes. Additionally, this method did not fully utilize the mined association relationships. To address this issue, we shifted from direct gene-level comparisons to a focus on association relationships, redesigning the Comparison View to visualize the relevance of different association relationships across selected regions. 
We also made adjustments to how we visualize relevance levels on this view. In the beginning, we used bar charts instead of donut charts. 
\sidecomment{R1C4\_12, R3C8}
\rui{We initially employed bar charts to visualize the mined association relationships, but the large and variable number of associations made it difficult to allocate sufficient space for each bar, hindering effective comparison. Moreover, bar charts were less effective at highlighting particularly relevant associations. To improve both space efficiency and perceptual salience, we adopted a donut-based design, where each complete ring corresponds to the upper quartile of the relevance distribution across the dataset. Associations extending beyond a single ring therefore indicate relevance levels exceeding this upper quartile threshold. This encoding allows experts to quickly identify highly relevant associations while maintaining a compact visual layout.}

\subsubsection{Verification View}
This view (\autoref{fig: teaser}-d) is designed to assist experts in biomarker construction and statistical validation. 
Experts can select two regions of cell populations from the Cell Exploration View for comparison, designating one as the positive group and the other as the negative group.
After selecting a set of genes as a biomarker for testing, the system automatically determines expression thresholds for each gene based on information entropy~\cite{Aevermann2021machine} and calculates the resulting classification accuracy under these thresholds \textit{\textbf{(T5)}}. 
Specifically, information entropy is defined as:  
\[
H(S) = - \sum_{i} p_i \log_2 p_i
\]
where \( p_i \) represents the proportion of samples belonging to a specific class within set \( S \). To determine an optimal threshold for each gene, candidate thresholds are evaluated based on their ability to partition the positive and negative groups while maximizing information gain, which is given by:  
\[
IG(T) = H(S) - \left( \frac{|S_1|}{|S|} H(S_1) + \frac{|S_2|}{|S|} H(S_2) \right)
\]
where \( S_1 \) and \( S_2 \) are the subsets formed by splitting \( S \) at threshold \( T \).
The threshold that yields the highest information gain is selected as the optimal cutoff point.  
Using these thresholds, the range of each gene within the positive group is highlighted visually in the Verification View. 
This view does not display exact thresholds, as only the relative high or low is meaningful for experts.  

\textbf{Interaction.} 
We provide an interactive panel that enables users to easily select and modify genes for validation \textit{\textbf{(T4)}}.
First, users can select suitable genes based on their importance levels across different association relationships.
To further enhance efficiency, we introduce a modification mechanism that enables users to refine and adjust previously defined biomarkers without having to restart the selection process but instead make quick adjustments based on their prior selections.



\section{Evaluation}
To validate whether our system can assist experts in completing the task of cell state discovery, we conducted three parts of the evaluation. First, we evaluated the performance of our MoE-based Mining Model. Second, we conducted expert interviews with six biologists who had not previously participated in our design process to understand their perspectives on the system's effectiveness. Finally, we presented a case based on a real-world dataset to demonstrate how our system specifically aids in the cell state discovery process.

\subsection{Model Evaluation} \label{model_evaluation}
To evaluate the effectiveness of our mining model, we conducted experiments on real-world datasets, using information entropy as a quantitative metric to assess the informativeness of the mined associations. First, we analyzed the impact of a key hyperparameter---the number of AI experts ($k$)---on the final results. By varying the number of AI experts, we examined its influence across multiple datasets (HCL~\footnote{http://bis.zju.edu.cn/HCL/contact.html}, EPI~\footnote{https://www.10xgenomics.com/products/epi-multiome}, and AQC~\footnote{https://db.cngb.org/nhpca/}) and reported the results for different expert configurations (\autoref{tab:expert_entropy}).
We can observe that as $k$ increases, the performance gradually stabilizes. This method allows us to determine the optimal value of $k$.

\begin{table}[htbp]
\centering
\caption{Information entropy across different numbers of AI experts.}
\label{tab:expert_entropy}
\begin{tabular}{c|ccccc}
\hline
\text{Dataset}  & \text{$k=4$} & \text{$k=6$} & \text{$k=8$} & \text{$k=10$} & \text{$k=12$}\\ 
\hline
HCL                              & 0.67               &0.69            &\textbf{0.73}&{0.72}&{0.72}  \\ 
EPI                              & 0.75               &0.74            &{0.79} &\textbf{0.80}&{0.79} \\ 
AQC                              & 0.71               &0.75            &{0.76} &\textbf{0.79}&{0.79} \\ 
\hline
\end{tabular}
\end{table}

Furthermore, our method inherently produces cell embeddings as intermediate outputs during the association mining process, offering a meaningful low-dimensional representation.
In this experiment, we compare our embeddings against four widely used dimensionality reduction methods for single-cell gene expression analysis (\ie, t-SNE, UMAP, IVIS, and DMT-EV~\cite{Zang2024DMT-EV}) to evaluate and demonstrate the effectiveness of our approach.
We also incorporated two additional datasets (\ie, MCA~\footnote{https://bis.zju.edu.cn/MCA} and GAST~\footnote{http://biogps.org/dataset/tag/gastric\%20carcinoma/}) for a more comprehensive evaluation. 
All of these single-cell datasets are accompanied by cell-state labels, providing pairs of gene expression profiles and their corresponding ground-truth categories. 

Firstly, to evaluate the ability of our method to preserve the global structure of the original data, we conducted a supervised classification experiment in which a support vector machine is trained on the low-dimensional embeddings generated by our approach.
The resulting test accuracy, reported in \autoref{tab:embedding_comparison}, reflects the ability of the embeddings to separate different cell states. High classification performance indicates that the embedding retains label-discriminative global structure, thereby supporting downstream supervised tasks such as cell state discovery. As shown in \autoref{tab:embedding_comparison}, our method achieves state-of-the-art performance on four datasets and attains an accuracy only marginally lower than the best-performing method on the remaining one dataset (MCA).

Secondly, we computed a clustering accuracy based on KNN label propagation (majority-vote among neighbors), which emphasizes local neighborhood purity and tests whether biologically similar cells are colocated in embedding space. 
As shown in \autoref{tab:embedding_comparison}, our embeddings achieve state-of-the-art clustering accuracy across all evaluated datasets, demonstrating superior preservation of the local manifold structure.

Thirdly, to evaluate intrinsic clustering structure independent of labels, we also report three classical internal validity indices: Calinski–Harabasz (CHI), which quantifies the ratio of between-cluster to within-cluster dispersion (higher is better); Davies–Bouldin (DBI), which measures average cluster similarity and thus penalizes poorly separated clusters (lower is better); and the Dunn index, which captures the smallest inter-cluster distance relative to the largest intra-cluster diameter (higher is better). As shown in \autoref{tab:embedding_comparison}, our method consistently achieves superior results across these internal metrics, indicating that the learned embeddings form clusters that are both highly compact and well separated—properties essential for accurate cell state discrimination and robust discovery of novel subpopulations.

\begin{table}[htbp]
    \centering
    \caption{Performance comparison on single-cell datasets.}
    \begin{tabular}{l|ccccc}
        \toprule
        \textbf{Dataset} & \textbf{t-SNE} & \textbf{UMAP} & \textbf{IVIS} &  \textbf{DMT-EV} & \textbf{Ours} \\
        \midrule
        \multicolumn{6}{c}{\textit{Classification Performance (Acc \( \uparrow \))}} \\ \midrule

        GAST                      & 61.4           & 49.5          & 63.5        & 75.0            & \textbf{80.4} \\
        HCL                       & 63.7           & 38.6          & 47.4           & 72.6            & \textbf{79.1} \\
        MCA                       & 44.7           & 38.2          & 69.7             & \textbf{77.4}   & 76.4          \\
        AQC                       & 74.6           & 67.8          & 51.2             & 70.6            & \textbf{82.9} \\
        EPI                       & 67.1           & 66.1          & 58.3             & 72.8            & \textbf{80.7} \\ \midrule
        \multicolumn{6}{c}{\textit{Clustering Performance (Acc \( \uparrow \))}} \\ \midrule
        GAST                      & 71.8           & 61.9          & 68.4           & 78.5            & \textbf{86.2} \\
        HCL                       & 68.2           & 43.2          & 50.0           & 74.3            & \textbf{80.3} \\
        MCA                       & 59.9           & 46.1          & 71.8             & 83.4            & \textbf{92.0} \\
        AQC                       & 82.3           & 75.0          & 50.4             & 80.4            & \textbf{88.2} \\
        EPI                       & 86.6           & 73.2          & 60.1             & 82.6            & \textbf{89.6} \\
        \midrule \multicolumn{6}{c}{\textit{Clustering Performance (CHI \( \uparrow \))}} \\ \midrule
        GAST                      & 420.3           & 380.7          & 450.5           & 520.8            & \textbf{560.1} \\
        HCL                       & 410.4           & 320.1          & 360.3           & 505.6            & \textbf{545.0} \\
        MCA                       & 400.2           & 350.5          & 480.7             & \textbf{560.3}            & 550.1 \\
        AQC                       & 460.6           & 410.2          & 380.4             & 495.1            & \textbf{575.4} \\
        EPI                       & 520.3           & 470.5          & 430.7             & 555.2            & \textbf{600.6} \\
        \midrule \multicolumn{6}{c}{\textit{Clustering Performance (DBI \( \downarrow \))}} \\ \midrule
        GAST                      & 0.86           & 0.94          & 0.82           & 0.75            & \textbf{0.68} \\
        HCL                       & 0.88           & 1.02          & 0.94           & 0.76            & \textbf{0.70} \\
        MCA                       & 0.92           & 1.01          & 0.84             & 0.76            & \textbf{0.70} \\
        AQC                       & 0.80           & 0.92          & 1.05             & 0.78            & \textbf{0.66} \\
        EPI                       & 0.74           & 0.88          & 0.96             & 0.71            & \textbf{0.63} \\
        \midrule \multicolumn{6}{c}{\textit{Clustering Performance (Dunn \( \uparrow \))}} \\ \midrule
        GAST                      & 0.42           & 0.39          & 0.45           & \textbf{0.51}            & 0.50 \\
        HCL                       & 0.44           & 0.35          & 0.38           & 0.52            & \textbf{0.56} \\
        MCA                       & 0.40           & 0.34          & 0.47             & 0.50            & \textbf{0.58} \\
        AQC                       & 0.49           & 0.40          & 0.32             & 0.51            & \textbf{0.60} \\
        EPI                       & 0.53           & 0.44          & 0.39             & 0.55            & \textbf{0.62} \\
        \bottomrule
    \end{tabular}
    \label{tab:embedding_comparison}
\end{table}

\subsection{Expert Interview}
To evaluate our system's effectiveness, we conducted individual interviews with five biologists (P1-P5) who specialize in cell state discovery.
\sidecomment{R2C3}
\rui{None of them had previously participated in the design of our system.}
P1 and P2 are PhD candidates with four years of experience, P3 and P4 are postdoctoral researchers with five and six years of experience, respectively, and P5 is a biology professor with over 20 years in the field.
The interview started with a 30-minute introduction, where we shared our research background and explained the features of our visual analytics system. 
After that, each participant had 15 minutes to explore the system on their own, with the opportunity to ask questions to help them get familiar with its functions.
\sidecomment{R2C3}
\rui{Next, the experts engaged in free exploration to identify cell states. Before the session, we asked them to provide datasets of interest, which we processed using our model. They then had half an hour to use our system and discover new cell states, during which we encouraged them to ``think aloud'' to help us capture detailed insights into their exploration process.}
After this exploration, each participant gave a brief 10-minute presentation.
Finally, we concluded the session with a 10-minute semi-structured interview to gather their feedback on the model's performance, visual design, user interactions, and suggestions for improvements.

\textbf{Model Performance.}
In our interviews, we asked participants to rate the association relationships mined by our model to evaluate its practical utility in exploration. We employed a 1-5 Likert scale, where higher scores indicate greater assistance from the model's mining results. The average score of the ratings is 4.4 (±0.5), suggesting that our model offers significant support across various exploration scenarios.
    For instance, P3 mentioned, \textit{``During my exploration, I discovered the interaction between MTCO2 and MTCO3, two mitochondrial genes that have been previously identified. This model helped me replicate earlier findings.''}
    P5 also found that CLCX2 and KRT15 seemed to distinguish a cell population, \textit{``This may be related to the cell states associated with tissue repair processes.}
    Finally, several participants (P1, P2, and P5) emphasized that each association can be explored from multiple aspects by enabling a flexible combination of different important gene types.
    For instance, P2 mentioned that the combination of KRT18, PIP, and CLDN4 is associated with epithelial cell function and tumorigenesis, while the combination of DCD, AZGP1, and KRT19 is linked to inflammation or metabolic states.
    These combinations can lead to potential biomarkers with diverse biological significance.

\textbf{System Workflow.}
Our system has received appreciation from five participants. Unlike traditional workflows, which primarily rely on biologists' subjective judgment and trial-and-error methods to select cell populations on two-dimensional cell representations, our system offers hints about the dataset in advance, providing a certain level of exploratory constraint.
P1 stated, \textit{``This constraint is a great thing; the exploration space is too vast.''}
P4 also mentioned, \textit{``For example, when I first examined the overview, I noticed there was an inconsistency in a cluster of cells that appeared to be grouped together, as different association relationships highlighted different regions. This served as a signal for my next steps, indicating that this cluster might be divided into several smaller groups.''}
Additionally, P3 emphasized two usage approaches of our system. First, when experts select a cell population, they can see which association relationships are strongly relevant.
Conversely, they can also first determine an association relationship they are very interested in and then identify which cells are strongly associated with it.
During P3's exploration, she discovered that a certain association relationship highlighted several genes related to immunity, which intrigued her very much. At that point, she chose the second exploration method.
However, when faced with the projection of cell representations, she began using the first method to understand some characteristics of the selected cells.
In the interview, P3 highlighted, \textit{``These two approaches can be flexibly switched between, making them very convenient during the exploration process.''}

\textbf{Design and Interaction.}
First, the overall design of our visual analytics system was considered relatively clear and simple during the expert interview.
For instance, P5 commented, \textit{``The central overview contains the most information, which I might need to study. However, considering that this is the most important view, I find it acceptable, and the learning cost is not high.''}
The designs of the other three views were regarded as very straightforward by our five participants.
Most importantly, the use of our metaphor was deemed very intuitive.
For example, P4 immediately asked upon seeing our system, \textit{``Is that a solar system in the middle?''}
During the exploration process, P1, P3, and P4 all noted that our design for visualizing genes changed their perspective on gene data presentation. Previously, they primarily relied on heatmaps to depict gene expression distribution, which often occupied significant space.
However, the moon visualization offered them a completely new approach. P1 remarked, \textit{``This visualization method is not only simple but also highly effective.''}
In addition, some of the interactions in our system were also appreciated by the participants.
For example, the Verification View allows users to make adjustments based on the currently selected gene types, which our participants found very convenient as it saved them the time of re-selecting.
Furthermore, P4 noted that our lasso selection operation was more effective than those in the systems she had used previously.
\textit{``Previously, our system only allowed us to select ranges from the x and y axes, resulting in a rectangular area for the selected cells. However, this selection method was not very precise. This lasso operation enables me to make much more nuanced selections.''}
Finally, our participants praised the system's ability to allow them to select colors for each association relationship by themselves.
For example, P5 mentioned, \textit{``For many, especially those with varying levels of color perception, customizing color choices enhances accessibility and usability---particularly when there are numerous extracted association relationships in this system.''}

\textbf{Suggestion.}
We also collected user suggestions on our visual analytics system.
Firstly, P1 proposed an approach to enhance the automation of gene selection within the system. He suggested that once a user selects two specific cell populations, the system could automatically recommend gene types that most effectively differentiate between these two groups of cells. These suggested gene types would then be displayed directly as ``moons'' adjacent to the various association relationships.
He noted, \textit{``Currently, users still need to manually select certain gene types based on their importance levels. This automated approach could reduce the effort required for users to make these selections.''}
However, he also acknowledged that our current manual selection method better supports experts' exploration interests.
In the future, we can explore ways to integrate both approaches to provide users with a more flexible experience.
Additionally, during the expert interview, P3 suggested that our system could benefit from incorporating a more intelligent component that can streamline the identification of potential biomarkers.
She highlighted that experts often brainstorm some potential biomarkers based on relevant literature.
In this context, she proposed whether our system could directly verify the specified biomarkers provided by experts based on the available gene expression data.
\textit{``This functionality could potentially integrate with the currently popular LLM-based multi-agent systems,''} P3 suggested, emphasizing the importance of leveraging advanced technologies to enhance user experience and improve research outcomes.

\subsection{Case Study}
With permission, we presented P2's exploration results from the expert interviews as a case study to showcase the effectiveness of our system.
Specifically, P2 explored our system using a real-world dataset~\cite{han2022cell}---the largest non-human primate single-cell transcriptomic atlas.
This dataset is crucial for advancing biologists' understanding of human biology, diseases, and aging, as it focuses on one of the species most closely related to humans---Macaca fascicularis. P2 specifically concentrated on epithelial cells, aiming to uncover novel cell states due to their essential roles in defense, secretion, and immune responses.

Our system first determined the optimal number of AI experts, which resulted in the selection of 10 AI experts. This led to the identification of 10 association relationships, each representing different key aspects of the data. P2 began examining the association relationships \textbf{(T1)} to understand the important genes involved in the AI Miner View through the bar charts (\autoref{fig: teaser}-a). For instance, she found that the \textbf{\textit{association relationship A}} primarily focused on the genes SYNM and LMO7, which are closely related to the structure of cells.
After reviewing each one, P2 became particularly interested in relationships B, C, and D.
\textit{``In the \textbf{association relationship B}, the most important genes seemed to be related to cell death, while the \textbf{association relationship C} was mainly linked to immune regulation and anti-inflammatory processes, and the \textbf{association relationship D} showed a strong connection to cancer-related genes.''}
These association relationships were closely related to the cell states of diseases that P2 aimed to explore.
She then clicked on these association relationships and assigned colors to them. After that, P2 started to observe the scatter plot in this view \textbf{(T1)} to understand their distribution (\autoref{fig: teaser}-a, the scatter plot).
Based on the highlighted visualization results (\autoref{fig: teaser}-a1), she noticed a region that was closely related to both association relationships C and D. This prompted her to investigate this area further.

\begin{figure}[ht]
    \centering
    \includegraphics[width=\linewidth]{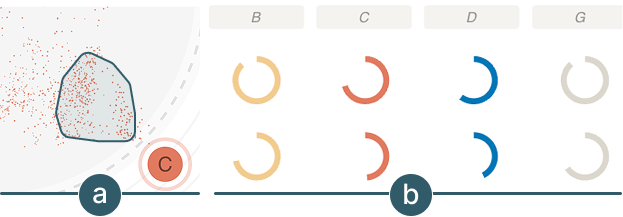}
    \vspace{-1.7em}
    \caption{(a) Examine the differences within a selected cell population through the \textit{\textbf{association relationship C}}. (b) Compare the two lasso-selected cell populations, with the top representing $region_1$ and the bottom representing $region_2$.}
    \label{fig: case1}
\end{figure}

P2 then began to explore the two-dimensional cell representations in the Cell Exploration View \textbf{(T2)}, focusing on the region (\autoref{fig: teaser}-b1) that she had previously found interesting. She observed a large number of cells in this area but was unsure whether it represented an individual cell state or if there might be further subdivisions within it.
To investigate, she hovered over the planet visualization representing the \textbf{\textit{association relationship C}}. This interaction caused the cell representation to change color based on its relevance with this association relationship---less relevant cells appeared more transparent.
Through observing the updated scatter plot (\autoref{fig: case1}-a), P2 found that this cluster of cell populations roughly exhibited two distinct distributions. One group of cell populations (\autoref{fig: case1}-a, the lasso-selected region) showed a higher relevance with the \textbf{\textit{association relationship C}}, while the other cell populations had a relatively lower relevance.
\textit{``It seems that although those cells are grouped together in the projection (\autoref{fig: teaser}-b1), based on the mining results, there might be some differences within it.''}
Therefore, P2 aimed to explore further to uncover the specific differences within this region and determine whether these differences would require subdividing those cell populations into distinct states.

\begin{figure}[ht]
    \centering
    \includegraphics[width=\linewidth]{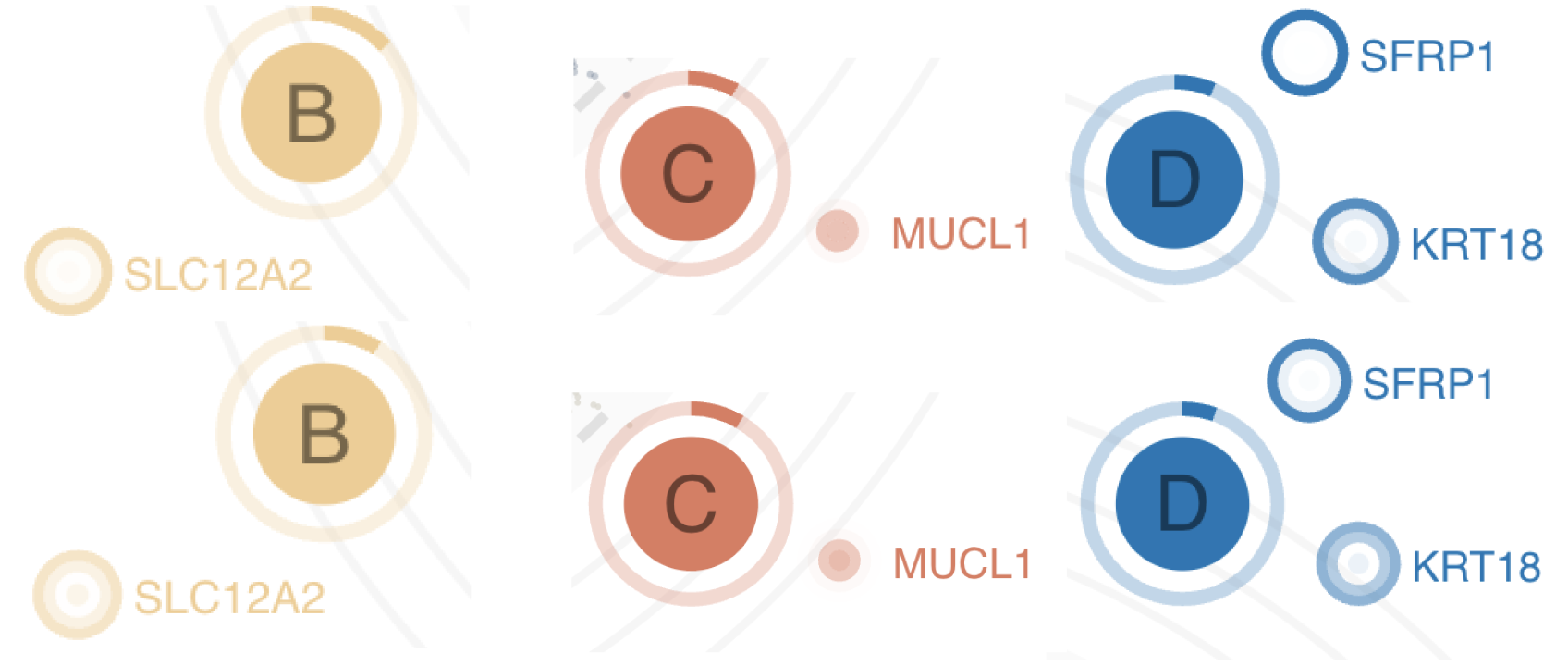}
    \vspace{-1.5em} \caption{In this figure, we present the gene types mentioned in our case study rather than displaying all the gene types selected by our expert. The top represents $region_1$, while the bottom represents $region_2$.}
    \label{fig: case2}
\end{figure}

Next, P2 started to compare the two regions of cell populations \textbf{(T3)} in the Comparison View. She first performed lasso operations on the scatter plot in the Cell Exploration View (\autoref{fig: teaser}-b, $region_1$ and $region_2$) and then hoped to comprehensively examine the differences in the Comparison View.
It was observed that the association relationships B, C, D, and G showed notable differences between the two lasso-selected regions (\autoref{fig: case1}-b, the comparison between the first and second row).
She then aimed to view the genes she was interested in to examine their distribution of expression levels. Therefore, she clicked on the planet visualizations representing the association relationships B, C, and D, which she believed were more closely related to diseases.
Then she selected gene types based on both her interests and their importance in each association relationship \textbf{(T4)}.
These selected genes appeared as moons orbiting their corresponding planets (\autoref{fig: case2}-b). Interestingly, when comparing the two regions, she noticed a remarkable distribution shift, with KRT18 and SLC12A2 exhibiting the most noticeable changes, as reflected in the variations of the corresponding moon visualizations.
Specifically, KRT18 is involved in maintaining cell structure and stability, while SCL12A2 plays a key role in ion transport and cell signaling.
P2 stated, \textit{``Previously, few studies have considered these two genes together as biomarkers, as although they may both be related to cancer, their functions are quite distinct. However, it now seems they may have the potential to work together in some way. I want to explore how effective they are as biomarkers when used in combination.''}

To assess the accuracy of this potential biomarker to distinguish the two selected regions \textbf{(T5)}, P2 added the two genes (\ie, KRT18 and SLC12A2) through the panel. She was surprised to find that the F1-score reached 0.83, with an accuracy of 0.72 (\autoref{fig: teaser}-d, the first card).
However, when testing these two genes individually, the F1-scores were both below 0.8.
\textit{``It appears that these two genes have a strong connection, which is an intriguing discovery. They seem to be more closely related to cancer than we initially believed.''}
After that, she also tested whether adding two other genes, MUCL1 and SFRP1, which are involved in processes like mucus production and cell signaling, and may also be linked to cancer, would improve the accuracy in distinguishing the two regions when paired with KRT18 and SLC12A2.
Through the experiment, the F1-score slightly increased to 0.84 (\autoref{fig: teaser}-d, the second card).
\textit{``This suggests that these two genes may provide some support. However, further research is needed, as these four genes together could be subtly linked and involved in changes in epithelial cell states.''}
In conclusion, P2 used our system to efficiently uncover differences within a cell population that appeared to be clustered together in a two-dimensional projection.
Additionally, she identified some unexpected genes that might form a biomarker driving this divergence. P2 stated, \textit{``The connections between these genes provided an unexpected insight and hold promising potential for identifying biomarkers and defining cell states in cancer research. This motivates me to further investigate the mechanisms of cooperation among these genes and define the characteristics of this cell state.''}

\section{Discussion}
In this section, we will first introduce three design lessons learned from our design system process.
Then, we will discuss the generalizability and scalability of our system. 
Finally, we will outline the limitations of our system and potential directions for future development.

\textbf{Design implications.}
Through our iterative design process, we identified three key lessons that informed our approach. The first is how to \textit{\textbf{integrate exploration objectives into visual analytics systems}}.
In the field of visualization, given that experts have varying prior knowledge and exploration objectives, different strategies have emerged to incorporate their expertise. For instance, one common approach is to leverage insights generated during the exploration process to facilitate active learning, allowing the system to adapt gradually to expert preferences. However, our work introduces another approach: modeling unclear exploration priors themselves as constraints. We also believe that in the future, this approach could be effectively combined with active learning for enhanced personalization.
The second one is that our design process also underscores \textit{\textbf{the importance of using metaphors that resonate with users’ existing knowledge}}. For instance, during the process of employing a solar system metaphor, we included familiar elements such as stars, planets, moons, and orbits, which users readily understood. However, when we introduced an asteroid belt (\ie, displaying all genes simultaneously around the association node) to enhance the theme of our metaphor in our previous design, many domain experts found it perplexing and questioned its effectiveness. This experience highlights a critical lesson: design does not require an overload of every conceivable component in metaphors. Instead, it should be carefully curated to align with users’ intuitive common knowledge. By concentrating on recognizable elements, we can create designs that enhance understanding and boost user engagement. 
The third one is the approach to \textit{\textbf{visualize gene expression data}}. Gene expression data, often consisting of hundreds or thousands of gene types, presents significant challenges for effective visualization. Many existing approaches attempt to leverage visualization formats that align with expert preferences, such as heat maps. 
Our approach highlights the effectiveness of building an intermediary bridge when visualizing such intricate information. 
By leveraging the mining association relationships, we are able to create a hierarchical structure that connects the cell representations to the gene expression distribution. This design prevents experts from being overwhelmed by the sheer volume of gene data at the outset. Additionally, it provides hints that guide experts toward potentially insightful gene types, facilitating a more targeted exploration of the data.
We also employ circle visualization to represent gene expression distributions, receiving positive feedback from our participants. This technique effectively conveys complex gene expression distribution while saving space, making it a practical choice for visualization. Although it may not be a familiar format for many domain experts at the beginning, its clear presentation and ability to simplify information finally highlight its value.

\textbf{Generalizability.}
Although our work primarily focuses on cell state discovery in the biomedical domain, it holds significant potential for application in other fields, such as market analysis, network security, and beyond. Specifically, our approach provides a novel perspective for uncovering hidden patterns within large and complex datasets. By combining mining algorithms with effective visualization designs, our method enables the discovery of previously unnoticed relationships in data.
For example, in market research, our MoE-based mining model can assist businesses in identifying potential customer segments or market niches without relying on pre-labeled customer data~\cite{ho2012customer}. This opens up the opportunity to uncover new consumer groups and behaviors that might not have been captured using traditional methods.

\textbf{Scalability.}
First, the number of association relationships mined by our model can impact the visualization results of our system. When the quantity is large, users may struggle to distinguish between them. To address this challenge, our system allows for flexible color assignment.
Specifically, users can customize the colors of different association relationships according to their needs. If users lose interest in certain relationships, they can revert them to the original gray color. This approach helps enhance visibility and clarity.
This also aligns with our observation that users tend to focus on specific association relationships during their exploration. In this context, our system allows them to dynamically adjust colors as needed for improved clarity.
Our model evaluation results (\autoref{model_evaluation}) also show that the three datasets we assessed reached performance stability with either 8 or 10 AI experts, which is not a large number. This finding suggests that our system's scalability is well-managed and poses minimal challenges.

Second, the number of cells in the dataset selected by biologists may affect our Cell Exploration View. When the number of cells is too large, it can lead to visual clutter, making it difficult for experts to examine where cell clusters exist. To alleviate this issue, when designing two-dimensional cell representations, we set the transparency for each cell point. 
Therefore, when more cells cluster together, even if there is an overlap in the scatter plot, our design can still help experts better perceive the clustering results in the Cell Exploration View. 
Furthermore, in our expert interviews, we found that the number of cells typically chosen to explore is between 5,000 and 10,000, as biologists usually select a cell category or several highly relevant categories of a dataset for their analysis.
This range is effectively supported by our system.
In the future, we can also consider providing zoom-in and zoom-out functions to help experts better view and analyze cell representation data.

\textbf{Limitations and future work.}
First, our current method for mining potential biomarkers does not adequately integrate biological knowledge. Some genes exhibit greater significance when considered together, yet our approach primarily relies on the judgment and selection of experts. To enhance the flexibility of these choices, we provide gene importance levels for each association relationship. This strategy partially addresses the challenges associated with presenting a fixed biomarker composition, which can complicate biological interpretation for researchers. In the future, we plan to incorporate multimodal machine-learning techniques to identify biomarkers with greater biological significance. 
Second, during the expert interviews, P1 suggested that we could make the selection of gene types more intelligent. 
This would require real-time monitoring of the experts' exploration process. We believe that integrating LLM-based multi-agent frameworks could assist in this endeavor, such as by introducing a monitoring agent.
Additionally, integrating gene databases is also a valuable and practical approach. These databases (\eg, Gene Ontology~\cite{Ontology2004Gene} and KEGG~\cite{kanehisa2002kegg}) offer rich biological context—such as gene functions, pathways, and known associations—which can be readily accessed using gene names. Incorporating such resources into our system would not only support basic annotation but also enable more intelligent recommendations of relevant gene sets or biomarkers during exploration. For example, when experts interact with specific associations or cell populations, the system could proactively suggest biologically related genes or pathways drawn from these databases.
\section{Conclusion}
In this work, we presented \systemname, a visual analytics system designed to support biologists in discovering new cell states and identifying their associated biomarkers. 
Specifically, our system is equipped with a MoE-based algorithm that can extract multiple association relationships, ensuring flexibility and relevance for biologists with varying exploration objectives. Additionally, we introduced a novel visualization interface inspired by the solar system metaphor, enabling experts to intuitively explore and refine the mined results.
To evaluate the effectiveness of our visual analytics system, we conducted a three-part evaluation: (1) assessing the performance of our MoE-based mining model, (2) interviewing five biologists outside our design process to gather their feedback, and (3) presenting a real-world case study.
In the future, we aim to explore how LLM-based multi-agent systems can further support this task and how to better integrate machine learning with visualization to create seamless workflows.
\bibliographystyle{ieeetr}
\bibliography{ref}

\begin{IEEEbiography}[{\includegraphics[width=1in,height=1.25in,clip,keepaspectratio]{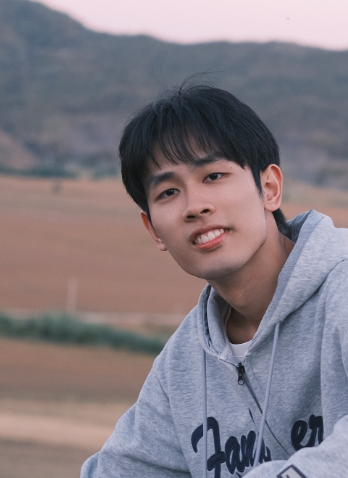}}]{Rui Sheng} is a Ph.D. candidate at HKUST (Hong Kong University of Science and Technology) VisLab under the supervision of Professor Huamin Qu. His research focuses on data visualization, human-AI collaboration, and decision-making, particularly in critical domains such as healthcare. He is dedicated to developing effective approaches for making informed decisions with AI assistance. More information: \url{https://dylansheng.github.io/}.
\end{IEEEbiography}
\begin{IEEEbiography}[{\includegraphics[width=1in,height=1.25in,clip,keepaspectratio]{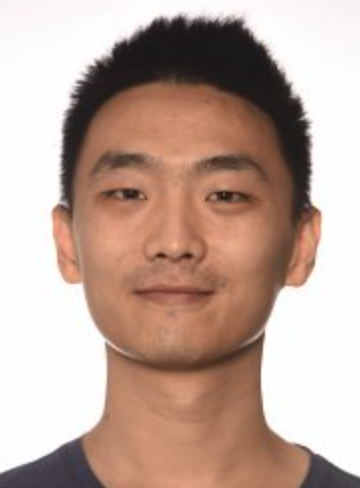}}]{Zelin Zang} has obtained his PhD degree at Westlake University, having previously completed his MEng degree at Zhejiang University of Technology. Specifically, he conducted his doctoral research at Westlake University under the supervision of the renowned Professor Stan Z. Li. His research interests focus on manifold learning, dimensionality reduction, and their applications in biological and image data analysis. His works explore the underlying structures of high-dimensional data and develop algorithms to enhance data processing efficiency and accuracy. 
\end{IEEEbiography}
\begin{IEEEbiography}[{\includegraphics[width=1in,height=1.25in,clip,keepaspectratio]{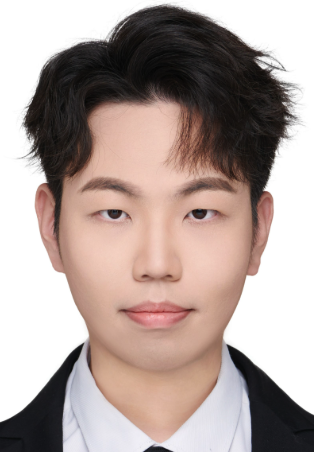}}]{Jiachen Wang} is a tenure-track assistant professor (ZJU 100-Young professor) at the Department of Sports Science, College of Education, Zhejiang University. He received his Ph.D. (supervised by Prof. Yingcai Wu) at the Department of Computer Science and Technology of Zhejiang University. Before that, he obtained his B.S. degree from Zhejiang University.
He majors in computational sports science which mainly focuses on knowledge discovery and visual analytics of sports data. More information: \url{https://www.wjc-vis.com/}.
\end{IEEEbiography}
\begin{IEEEbiography}[{\includegraphics[width=1in,height=1.25in,clip,keepaspectratio]{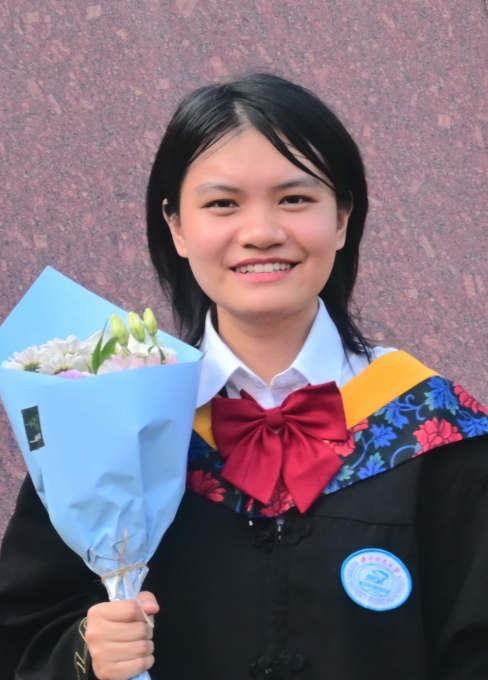}}]{Yan Luo} is a Mphil student in the Vislab affiliated with the Hong Kong University of Science and Technology(HKUST) under the guidance of prof. Huamin Qu. Before that, she received her Bachelor’s Degree from Huazhong University of Science and Technology in Wuhan. Her interest lies in Data Visualization and Human-Computer Interaction(HCI). She has experience in developing visual analytics systems. More information: \url{https://windyan233.github.io/}.
\end{IEEEbiography}
\begin{IEEEbiography}[{\includegraphics[width=1in,height=1.25in,clip,keepaspectratio]{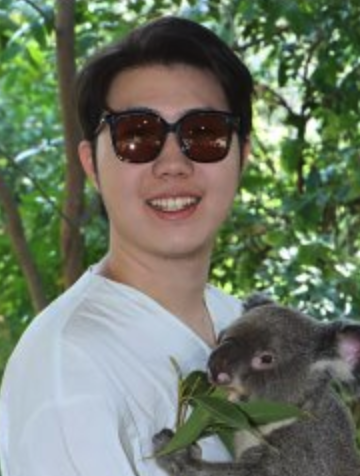}}]{Zixin Chen} is currently a Ph.D. candidate from Hong Kong University of Science and Technology. His research centers on Data Visualization and Human-AI Collaboration, with a particular emphasis on leveraging them in LLM for Education. He designs interactive tools that support teaching and pedagogical decision-making, personalize learning experiences, and makes LLM behavior more trustworthy for both learners and educators. In parallel, he explores the broader applicability of LLM-driven models and systems in other significant domains (AI4Health and AI4Science). More information: \url{https://cinderd.github.io/}.
\end{IEEEbiography}
\begin{IEEEbiography}[{\includegraphics[width=1in,height=1.25in,clip,keepaspectratio]{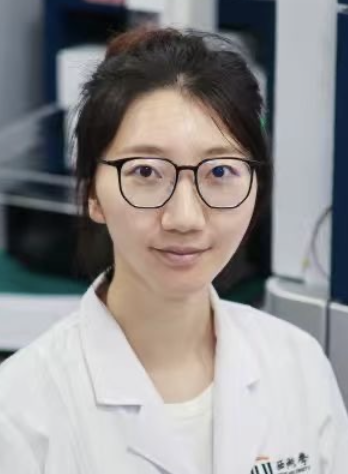}}]{Yan Zhou} received her bachelor’s degree in biology from the University of Manchester and Shandong University, followed by a master’s degree in computational biology from University College London. She joined Prof. Tiannan Guo's lab as a PhD student in 2021, with research focusing primarily on clinical proteomicså.
\end{IEEEbiography}
\begin{IEEEbiography}[{\includegraphics[width=1in,height=1.25in,clip,keepaspectratio]{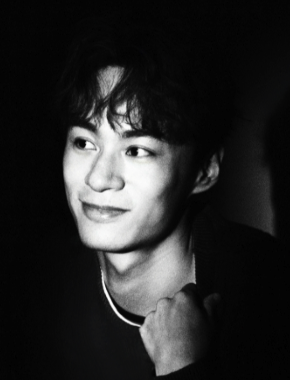}}]{Shaolun Ruan} is a Ph.D. candidate of Computer Science at Singapore Management University, under the supervision of Professor Yong WANG and Professor Jiannan LI. His research interests include Data Visualization and Human-Computer Interaction. His work focuses on developing human-centered computing tools to address complex scientific problems, facilitating the process of explainability and data-driven decision-making. More information: \url{https://shaolun-ruan.com/}.
\end{IEEEbiography}
\begin{IEEEbiography}[{\includegraphics[width=1in,height=1.25in,clip,keepaspectratio]{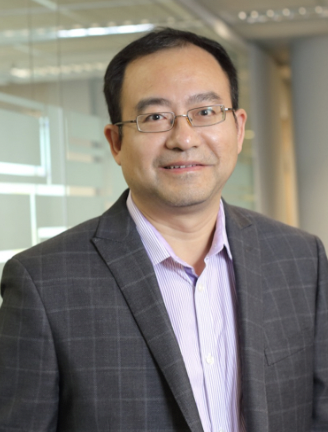}}]{Huamin Qu} is the dean of the Academy of Interdisciplinary Studies, head of the Division of Emerging Interdisciplinary Areas, and a chair professor in the Department of Computer Science and Engineering at HKUST. He obtained a BS in Mathematics from Xi’an Jiaotong University, China, an MS, and a PhD in Computer Science from the Stony Brook University. His main research interests are in visualization and human-computer interaction, with focuses on urban informatics, social network analysis, Elearning, text visualization, and explainable artificial intelligence More information: \url{http://huamin.org/}.
\end{IEEEbiography}
\end{document}